\documentclass[twocolumn,aps,superscriptaddress]{revtex4}
\usepackage{amssymb}
\usepackage{amsmath}
\usepackage{graphicx}
\usepackage[normalem]{ulem}
\usepackage[dvips]{color}
\usepackage{appendix}

\begin{document}

\title{Hadronization using the Wigner function approach for a multiphase transport model}
\author{Feng-Tao Wang}
\affiliation{Shanghai Institute of Applied Physics, Chinese Academy
of Sciences, Shanghai 201800, China}
\affiliation{University of Chinese Academy of Sciences, Beijing 100049, China}
\author{Jun Xu\footnote{corresponding author: xujun@zjlab.org.cn}}
\affiliation{Shanghai Advanced Research Institute, Chinese Academy of Sciences,
Shanghai 201210, China}
\affiliation{Shanghai Institute of Applied Physics, Chinese Academy
of Sciences, Shanghai 201800, China}

\date{\today}

\begin{abstract}
In the string melting version of a multiphase transport model, the hadronization algorithm has been improved by favoring parton combinations close in not only coordinate space but also momentum space. Formation probabilities of mesons, baryons, and antibaryons during hadronization are determined by their corresponding Wigner functions for the valence parton combinations with no free parameters, and the net baryon, electric, and strangeness charges are conserved during quark coalescence. Effects of the hadronization on anisotropic flows, collision energy dependence of the proton directed flow, relative particle yield ratios, as well as di-hadron correlations in relativistic heavy-ion collisions at the energies ranging from RHIC beam energy scan to LHC have been extensively discussed.
\end{abstract}

\maketitle

\section{INTRODUCTION}
\label{introductionhan}

One of the main goals of relativistic heavy-ion collision experiments is to study the properties of the quark-gluon plasma (QGP), which can be produced at extremely high temperatures and densities at the Relativistic Heavy Ion Collider (RHIC) and the Large Hadron Collider (LHC). In order to search for the signals of the critical point of the quark-hadron phase transition, RHIC beam energy scan (BES) program has also been proceeding collisions at relatively lower energies. As a relativistic heavy-ion collision lasts for a very short time, our understandings of its dynamics often rely on transport models~\cite{Bas98,Xu05,Lin05,Cas09}, hydrodynamic models~\cite{Huo01,Bet09,Sch11,Boz12}, or hybrid models~\cite{Pet08,Wer10,Son12}. Since the experimental observables are final hadrons, the hadronization mechanism converting partons to hadrons in these dynamic models is extremely important for us to understand the properties of the QGP indirectly.

There are various treatments describing the quark-hadron phase transition used in the dynamic models for relativistic heavy-ion collisions. The quark coalescence model has been used to successfully describe the hadron yield and elliptic flow~\cite{Gre03,Gres03,Fri03,Fris03,Hwa03}. For the hadronization of high-momentum partons, the fragmentation mechanism becomes the dominating one (see, e.g., Ref.~\cite{And83}). In the hydrodynamic models, the Cooper-Frye mechanism~\cite{Coo74} that produces locally thermalized hadrons is generally employed. The dynamical quasiparticle approach, that forms ground-state mesons or octet baryons from sequential decays of the $q\bar{q}$ or the $qqq$ resonants, has also been used in transport model simulations~\cite{Cas08}. These different hadronization treatments are expected to affect the simulation results that are used to extract the properties of the QGP.

Collective flows are important experimental observables characterizing the properties of the QGP. The number-of-constituent-quark (NCQ) scaling law of anisotropic flows $v_n$, i.e., $v_{n} / n_{q} \sim p_{T} / n_{q}$~\cite{STAR04,STAR05a,STAR05b,PHENIX07a,PHENIX07b,STAR08}, with $p_{T}$ being the transverse momentum, and $n_q$ being the number of constituent quarks, was considered as an evidence of the QGP formation, and can be well explained by the naive coalescence model which assumes that hadrons are formed by constituent quarks with the same momentum~\cite{Mol03,Kol04}. It was also found that different orders of anisotropic flows may follow the scaling relation $v_n \sim v_2^{n/2}$~\cite{ATLAS12}, studying which is helpful in understand the behavior of the initial eccentricities~\cite{Lac10,Lac11b}, the viscous property of the QGP~\cite{Gar12}, and the acoustic nature of anisotropic flows~\cite{Lac11a}. In our previous study~\cite{Zha16}, we found that the hadronization mechanism has large impacts on the scaling relation of anisotropic flows. On the other hand, the slope of the directed flow $v_{1}$ at midrapidities ($y \sim 0$) has been shown to be sensitive to the equation of state of the produced matter from both hydrodynamic~\cite{Sto05} and transport~\cite{Bas98} model studies. These studies predicted that the slope of $v_{1}(y)$ near the midrapidity region ($d v_{1} / dy|_{y=0}$) changes sign from positive to negative~\cite{Bra00,Cse99,Sne00} with increasing collision energies, and this was indeed observed from the RHIC-BES program~\cite{Ada14,Ada18}, where the slope of the proton directed flow at midrapidities changes sign from positive to negative between $\sqrt{s_{NN}}=7.7$ GeV and 11.5 GeV. Considerable efforts~\cite{Ste14,Kon14,Nar16} have been devoted to explaining the directed flow results, but so far no approach can reproduce the experimental directed flow at various collision energies very satisfactorily. In our previous studies~\cite{Guo17,Guo18}, it was found that the hadronization mechanism affects significantly the proton directed flow.

The relative particle yield ratios in relativistic heavy-ion collisions are also important observables, which can be used to extract the temperature $T$ and the chemical potential $\mu$ at chemical freeze-out based on the statistical model~\cite{sta,Cel06}. This is of special importance in the RHIC-BES program, where values of $(T, \mu)$ extracted at different collision energies leave trajectories in the QCD phase diagram~\cite{STAR17}, from which the information of the QCD critical point can be obtained. In transport simulations, the relative hadron yield ratios are not only determined by the initial quark species but also affected by the hadronization mechanism.

Recently, the ALICE Collaboration have measured the di-hadron correlations in p+p collisions at $\sqrt{s_{NN}}=7$ TeV~\cite{Ada17a}. At the near side, it was found that the meson-meson correlation exhibits an expected peak, while the baryon-baryon and antibaryon-antibaryon correlations show an anti-correlation structure. It was argued that this phenomenon can be related to the baryon production mechanisms in the fragmentation process~\cite{Ada17a}. It was later found that an improved hadronization mechanism indeed helps to  reproduce the anti-correlation structure at the near side of the baryon-baryon and antibaryon-antibaryon correlations~\cite{Zha18}.

In the present study, we attempt to improve the hadronization algorithm in the string melting version of a multiphase transport (AMPT) model~\cite{Lin05}, and investigate the effect on anisotropic flows, energy dependence of the proton directed flow, relative particle yield ratios, as well as di-hadron correlations in relativistic heavy-ion collisions. Different from the previous spatial coalescence approach, parton combinations with a small relative distance in coordinate and momentum space now have a larger probability to coalesce into hadrons, while the corresponding Wigner functions describing the coalescence probability are determined by properties of pions and protons as representative mesons and baryons. We found that the improved coalescence algorithm preserves better information of the partonic dynamics in transport simulations of relativistic heavy-ion collisions.

The rest part of the paper is organized as follows. Section~\ref{theorywang} gives a brief review of the structure of the AMPT model, and describes in detail the improved coalescence algorithm. Section~\ref{resultwang} compares results of anisotropic flows, energy dependence of the directed flow, relative particle yield ratios, and di-hadron correlations from the original and the improved AMPT model. A summary is given in Sec.~\ref{summary}.

\section{Framework of AMPT model with improved hadronization algorithm}
\label{theorywang}

The string melting version of the AMPT model~\cite{Lin05}, which is used in the present study, mainly consists of four parts: the initial condition generated by the heavy-ion jet interaction generator (HIJING) model~\cite{Xnw91}, the partonic evolution described by Zhang's parton cascade (ZPC) model~\cite{Zha98}, the hadronization process, and the hadronic evolution described by a relativistic transport (ART) model~\cite{Li95}. The initial momentum distribution of partons is obtained by melting hadrons produced by elastic and inelastic scatterings of participant nucleons in HIJING. The positions of partons in the transverse plane are set as the same as colliding nucleons that produce these partons, while the longitudinal coordinates of initial partons are sampled uniformly within $(-lm_N/\sqrt{s_{NN}}, lm_N/\sqrt{s_{NN}})$, where $l$ is the size scale of colliding nuclei, and $m_N=0.938$ GeV is the nucleon mass. The partonic interaction in the ZPC model is described by two-body elastic scatterings with the cross section given by $\sigma \approx {9 \pi \alpha_{s}^{2}}/{2 \mu^{2}}$, where the parton screening mass $\mu$ is fixed at 3.2264 $\mathrm{fm}^{-1}$ and the strong coupling constant $\alpha_{s}$ is set to be 0.33, 0.38, and 0.47 for the cross section of 1.5 mb, 2 mb, and 3 mb, respectively. The partons freeze out kinetically after their last scatterings at different times, and afterwards a coalescence algorithm is used to combine partons into hadrons to be detailed in the following. The hadronic evolution is described by the ART model with various of elastic, inelastic, and decay channels, until all hadrons freeze out kinetically.

In the original AMPT model, a pair of quark and antiquark, which are nearest in coordinate space in their center-of-mass (C.M.) frame, can form a meson. Similarly, three nearest quarks (antiquarks) can form a baryon (antibaryon). The species of the hadron after coalescence is determined by the flavors and the invariant mass of its constituent partons. The 3-momentum is conserved in the coalescence procedure, and the formation time of the hadron is also related to the freeze-out times of its constituent partons. In addition, the numbers of mesons, baryons, and antibaryons after coalescence are the same as their initial numbers before hadrons from HIJING are melted into partons.

The original coalescence algorithm has the ambiguity of the coalescence ordering for mesons or baryons (antibaryons), and there is no physics foundation of fixing the numbers of mesons, baryons, and antibaryons as the initial ones. Recently, He and Lin~\cite{He17} made an improvement to the above algorithm by introducing a parameter controlling the preference of partons to coalesce into mesons or baryons (antibaryons), helping to better reproduce the relative particle yield ratio and removing the ambiguity of the coalescence ordering. Since all partons are forced to be used up, the net baryon, electric, and strangeness charges are always conserved as in the original hadronization. However, so far the hadronization algorithm is still restricted to the parton combinations close in coordinate space.

The hadronization algorithm can be further improved based on the dynamical coalescence approach~\cite{Gre03,Gres03}, where the coalescence probability is described by the hadron Wigner function depending on the relative coordinates of constituent partons obtained from Jacobi transformation (see, e.g., Ref.~\cite{Sun17}). For a pair of quark and antiquark to form a meson, the corresponding Wigner function is expressed as
\begin{equation}
f_{M}(\boldsymbol{\rho}, \boldsymbol{k}_{\rho})=8 g_{M} \exp \left(-\frac{\boldsymbol{\rho}^{2}}{\sigma_{\rho}^{2}}-\boldsymbol{k}_{\rho}^{2} \sigma_{\rho}^{2}\right).
\end{equation}
In the above, $g_{M}$ is the statistical factor for mesons and set to be 1/36 for pions, and
\begin{equation}
\begin{aligned} \rho &=\frac{1}{\sqrt{2}}(\boldsymbol{r}_{1}-\boldsymbol{r}_{2}), \\ \boldsymbol{k}_{\rho} &=\sqrt{2} \frac{m_{2} \boldsymbol{k}_{1}-m_{1} \boldsymbol{k}_{2}}{m_{1}+m_{2}} \end{aligned}
\end{equation}
are the relative distance in coordinate and momentum space in the C.M. frame of the quark-antiquark system, with $\boldsymbol{k}_{i}$, $\boldsymbol{r}_{i}$, and $m_{i}$ being the momentum, coordinate, and mass of the $i$th parton, respectively. The width parameter $\sigma_{\rho}$ is determined by the root-mean-square (RMS) radius $\sqrt{\langle r_{M}^{2}\rangle}$ of the meson through the relation
\begin{equation}
\begin{aligned}
\langle r_{M}^{2}\rangle &=\frac{3}{2} \frac{m_{1}^{2}+m_{2}^{2}}{(m_{1}+m_{2})^{2}} \sigma_{\rho}^{2} \\ &=\frac{3}{4} \frac{m_{1}^{2}+m_{2}^{2}}{m_{1} m_{2}(m_{1}+m_{2}) \omega}. \end{aligned}
\end{equation}
In the second line of the above expression we used the relation $\sigma_{\rho}$ = 1$/ \sqrt{\mu_{1} \omega}$  in terms of the reduced mass $\mu_{1}=2(1 / m_{1}+1 / m_{2})^{-1}$ and the frequency parameter $\omega$.
Similarly, the probability for three quarks (antiquarks) to form a baryon (antibaryon) can also be given by the corresponding Wigner function as
\begin{equation}
\begin{array}{l}{f_{B}(\rho, \lambda, k_{\rho}, k_{\lambda})}
{=8^{2} g_{B} \exp\left(-\frac{\rho^{2}}{\sigma_{\rho}^{2}}-\frac{\lambda^{2}}{\sigma_{\lambda}^{2}}-k_{\rho}^{2} \sigma_{\rho}^{2}-k_{\lambda}^{2} \sigma_{\lambda}^{2}\right)},
\end{array}
\end{equation}
where $g_{B}$ is the statistical factor for baryons and set to be 1/108 for protons, the Gaussian width parameter can be expressed as $\sigma_{\lambda}$ = 1$/ \sqrt{\mu_{2} \omega}$ with $\mu_{2}$ = $(3 / 2)[1 /(m_{1}+m_{2})+1 / m_{3}]^{-1}$, and
\begin{equation}
\begin{aligned}
\boldsymbol{\lambda} &=\sqrt{\frac{2}{3}}\left(\frac{m_{1} \boldsymbol{r}_{1}+m_{2} \boldsymbol{r}_{2}}{m_{1}+m_{2}}-\boldsymbol{r}_{3}\right), \\
\boldsymbol{k}_{\lambda} &=\sqrt{\frac{3}{2}} \frac{m_{3}(\boldsymbol{k}_{1}+\boldsymbol{k}_{2})-(m_{1}+m_{2}) \boldsymbol{k}_{3}}{m_{1}+m_{2}+m_{3}}
\end{aligned}
\end{equation}
are the relative distance in coordinate and momentum space in the C.M. frame of the three-parton system, respectively. Although the value of an individual Wigner function for baryons (antibaryons) can be different by generally $10-20\%$ depending on the ordering of constituent partons, the statistical results are expected to be independent of the ordering from a smooth parton distribution. The Gaussian width parameter is determined by the RMS radius $\sqrt{\langle r_{B}^{2}\rangle}$ of the baryon (antibaryon) through the relation
\begin{equation}
{\langle r_{B}^{2}\rangle=} \\  {\frac{1}{2}\frac{m_{1}^{2}(m_{2}+m_{3})+m_{2}^{2}(m_{1}+m_{3})+m_{3}^{2}(m_{1}+m_{2})}{(m_{1}+m_{2}+m_{3})m_{1} m_{3} \omega}}.
\end{equation}
As representatives for mesons and baryons, the RMS radii of 0.61 fm for $\pi^{+}$ and 0.877 fm for protons~\cite{Ber12} are used to determining the widths of the corresponding Wigner functions, so there is no free parameter in the new coalescence algorithm.

Instead of choosing combinations of nearest partons in coordinate space, we now choose combinations of partons with the largest Wigner function in their C.M. frame. Practically, based on the phase-space information of freeze-out partons, we search for all possible combinations of quark-antiquark pairs, three quarks, or three antiquarks, which can potentially form mesons, baryons, and antibaryons, respectively, and the maximum values for each kind of combinations are $f_M$, $f_B$, and $f_{\bar{B}}$, whose values decide which kind of hadrons is first formed from which two or three partons, i.e.,
\begin{gather}\label{originalLag}
f_{B} = {\rm Max}\{f_{M},f_{B},f_{\bar{B}}\} : \rm form\; a\; baryon; \notag \\
f_{\bar{B}} = {\rm Max}\{f_{M},f_{B},f_{\bar{B}}\} : \rm form\; an\; antibaryon; \notag \\
f_{M} = {\rm Max}\{f_{M},f_{B},f_{\bar{B}}\} : \rm form\; a\; meson.
\end{gather}
Then the combinations for the rest of partons are considered, and this process lasts until all partons are used up. It is thus see that the preference to form a meson or a baryon (antibaryon) is entirely determined by the distance among the constituent partons in phase space, and there is no ambiguity of the coalescence ordering or number constraints. In addition to combining partons close in coordinate space to preserve the collision geometry, the improved coalescence algorithm combines partons with the similar momentum in a more physical picture. This may help to preserve the information of the dynamics in the partonic phase after hadronization. Furthermore, the 3-momentum conservation condition is satisfied in the improved coalescence algorithm, and the hadron species are determined in the same way as in the original AMPT model.

\section{Results and discussions}
\label{resultwang}

In this section, we investigate the effects of the hadronization mechanism on anisotropic flows at the top RHIC energy and the LHC energy, directed flows at RHIC-BES energies, relative particle yield ratios at all collision energies, and di-hadron correlations in p+p collisions at the LHC energy. Results from the original AMPT model with the spatial coalescence, the improved AMPT model with the quark coalescence using the Wigner function approach, and the available experimental data are compared. It is found that the improved AMPT model has the similar multiplicity distribution as the original AMPT model, so the Lund string fragmentation parameters are set to be the same as those used in our previous studies, i.e., $a$ = 0.5 and $b$ = 0.9 $\mathrm{GeV}^{-2}$ for Au+Au collisions at 200 GeV~\cite{Xu11} and Pb+Pb collisions at 2.76 TeV~\cite{Xu11s}, and $a$ = 2.2 and $b$ = 0.5 $\mathrm{GeV}^{-2}$ for Au+Au collisions from 7.7 GeV to 39 TeV~\cite{Zha17}. The relation between the impact parameter b and the centrality $c$ is taken from the empirical relation $c=$ $\pi {\rm b}^{2} / \sigma_{\rm in}$~\cite{Bro02,Xu11,Xu11s}, where $\sigma_{\text { in }}$ is the total inelastic cross section at the corresponding collision energy. The parton scattering cross section is fitted so that the final charged particle elliptic flows from the AMPT model reproduce the experimental data.

\subsection{Relative distance among partons in hadrons}

\begin{figure}[ht]
	\includegraphics[scale=0.15]{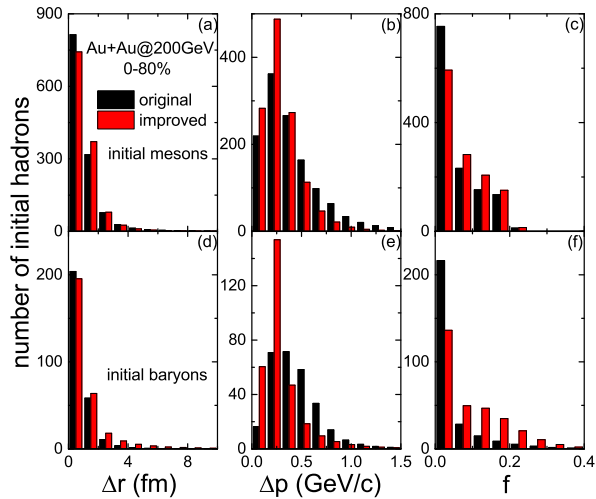}
	\caption{(Color online) Histograms of the relative distance in coordinate space ($\Delta r$) [(a), (d)], in momentum space ($\Delta p$) [(b), (e)], and the value of the Wigner function $f$ [(c), (f)] among valence partons in their C.M. frame for the initial mesons (upper) and the initial baryons (lower) in the hadronization process in minibias Au+Au collisions at $\sqrt{s_{NN}}=200$ GeV. Results from the original (black) and the improved (red) AMPT model are compared.} \label{rpf}
\end{figure}

Before discussing the physics results, we first check with the distribution of the distance in coordinate space ($\Delta r$) and that in momentum space ($\Delta p$) among valence partons for the initial hadrons after hadronization. For initial mesons, $\Delta r$ and $\Delta p$ are those between their constituent quarks and antiquarks in their C.M. frame. For initial baryons, $\Delta r$ and $\Delta p$ are the average distances in coordinate and momentum space among the three constituent quarks. These initial hadrons includes resonances with their masses determined by the invariant masses of constituent partons. Figure~\ref{rpf} displays the corresponding histograms for minibias Au+Au collisions at $\sqrt{s_{NN}}=200$ GeV as an example. Compared with the histograms from the original AMPT model, the combinations of valence partons for both mesons and baryons from the improved AMPT model have on average slightly larger $\Delta r$ but smaller $\Delta p$. It is thus seen that the improved AMPT model scarifies slightly the $\Delta r$ distribution in order to achieve a better $\Delta p$ distribution, and gives essentially on average larger values of the Wigner function $f$ compared with the original AMPT model, as shown in the right column of Fig.~\ref{rpf}. The scenario with parton combinations that have overall larger values of the coalescence probability is more physical. On the other hand, the coalescence for partons close in momentum space helps to enlarge the hadron transverse momenta and thus stiffen the transverse momentum spectra, which are generally soft in the AMPT model due to massless partons less driven by the radial flow. Although the transverse momentum spectra from the improved AMPT model are found to be still softer than the experimental data, they can be stiffer with a very small Lund string fragmentation parameter $b$, such as that used in Ref.~\cite{He17}.

\subsection{Effect of coalescence algorithm on hadron elliptic flow}

\begin{figure}[ht]
	\includegraphics[scale=0.3]{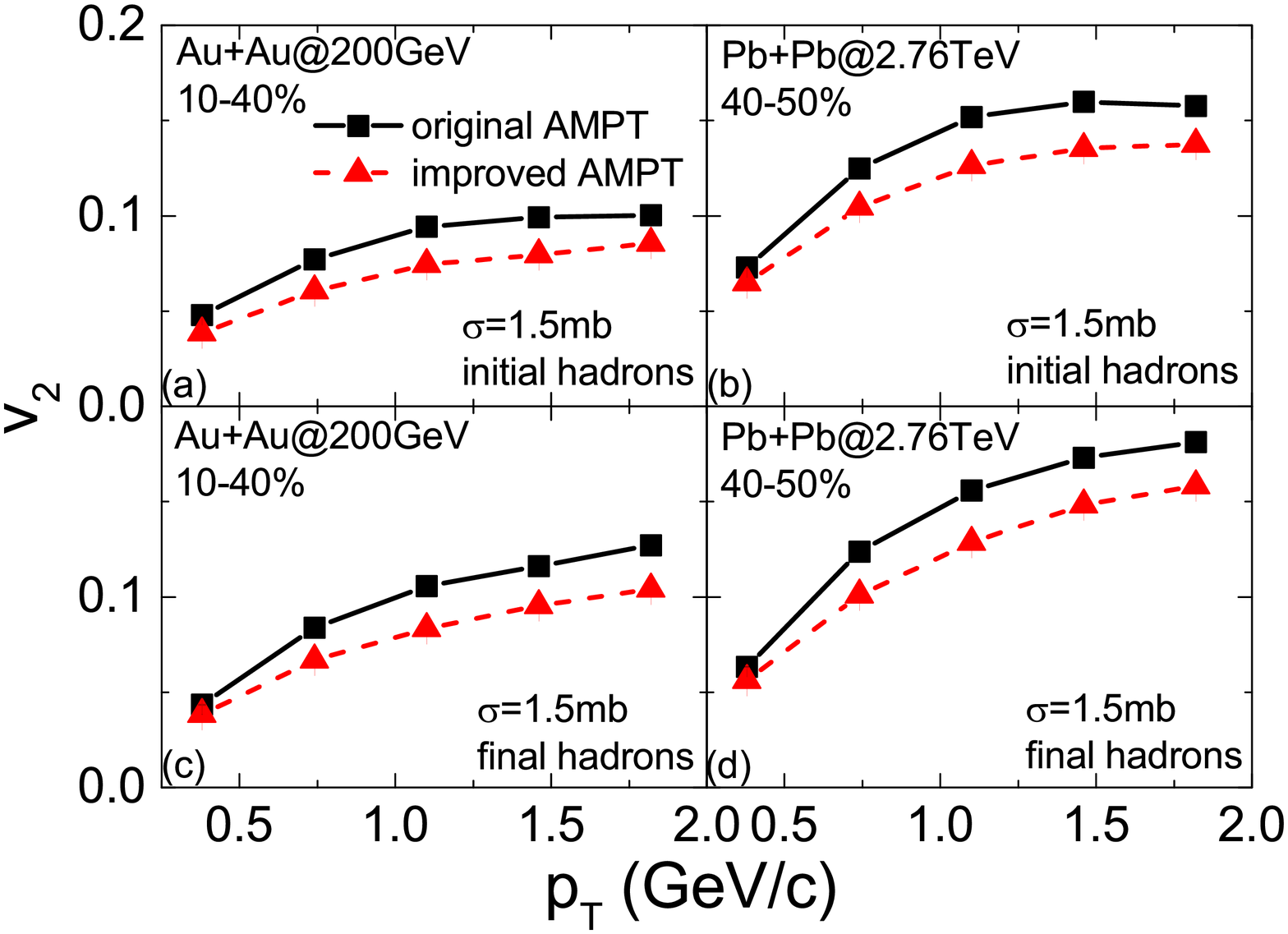}
	\caption{(Color online) Comparison of the initial and final hadron elliptic flows ($v_{2}$) at midpseudorapidities as functions of the transverse momentum ($p_{T}$) in midcentral ($10 \%-40 \%$) Au+Au collisions at $\sqrt{s_{NN}}=200$ GeV and midcentral ($40 \%-50 \%$) Pb+Pb collisions at $\sqrt{s_{NN}}=2.76$ TeV from the original and the improved AMPT model using the same parton scattering cross section of $\sigma=1.5$ mb.}
	\label{v2compare}
\end{figure}

Despite the fact that the elliptic flow from the AMPT model can be affected by many sources (see, e.g., Ref.~\cite{Mol19}), it has been used to extract the specific shear viscosity of the quark medium at RHIC and LHC energies~\cite{Xu11,Xu11s}. However, the elliptic flow of hadrons is not only dominated by the parton scattering cross section, but is also affected significantly by the hadronization algorithm. Figure~\ref{v2compare} compares the transverse momentum dependence of $v_2$ at top RHIC and LHC energies from the original and the improved AMPT model using the same parton scattering cross section but different hadronization algorithms. From the same parton freeze-out phase-space distribution, it is seen that the initial hadron $v_2$ right after hadronization is generally larger from the original AMPT model using a spatial coalescence, compared with the improved AMPT model favoring parton combinations close in phase space. This is possible if the momentum spectrum of freeze-out partons in the reaction plane is stiffer than that out of the reaction plane as observed in ZPC/AMPT, thus hadrons even from the spatial coalescence keep the strong in-plane motion. The improved coalescence algorithm, which is closer to the naive coalescence scenario, leads to the comoving of constituent partons, while the motions of formed hadrons are actually more randomized, leading to a weaker hadron $v_2$. It is found that $v_2$ is even smaller from the usual dynamical coalescence method based on the same freeze-out parton phase-space distribution. The eccentricity of the initial hadronic phase is found to be slightly larger in the original AMPT model than in the improved AMPT model, since the former preserves better the spatial distribution during the hadronization, as can be expected from the $\Delta r$ histogram in Fig.~\ref{rpf}. The final hadron $v_2$ after the hadronic evolution is also seen to be larger from the original AMPT model than from the improved AMPT model.

\begin{figure}[ht]
	\includegraphics[scale=0.3]{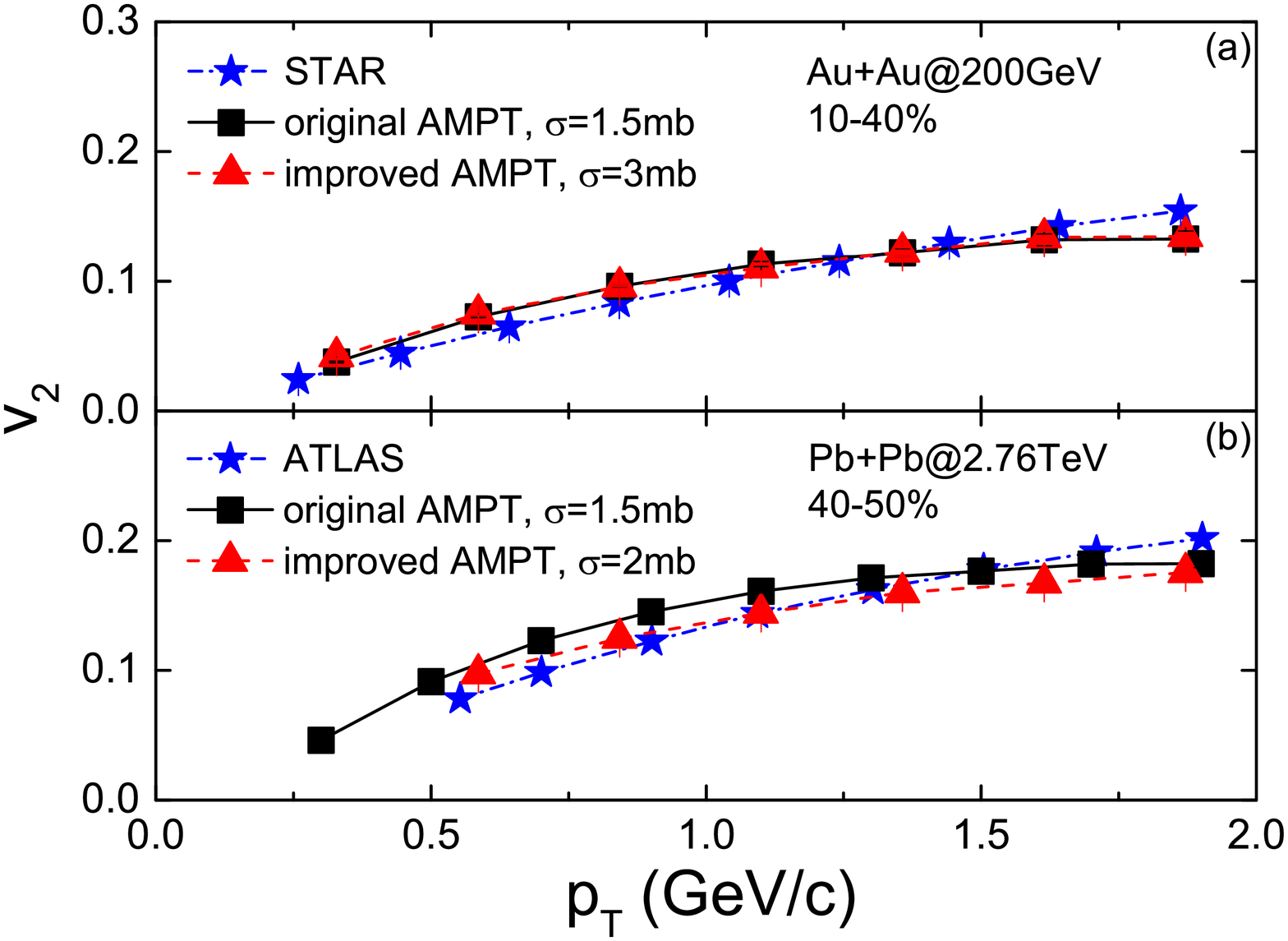}
	\caption{(Color online)  Transverse momentum dependence of the charged hadron elliptic flow at midpseudorapidities in midcentral ($10 \%-40 \%$) Au+Au collisions at $\sqrt{s_{NN}}=200$ GeV (a) and midcentral ($40 \%-50 \%$) Pb+Pb collisions at $\sqrt{s_{NN}}=2.76$ TeV (b). Results from the original and the improved AMPT model using different parton scattering cross sections are compared with the experimental data from the STAR~\cite{Abe08} and the ATLAS~\cite{Aad12} Collaborations.}
	\label{v2pt}
\end{figure}

In order to reproduce the transverse momentum dependence of the elliptic flow at top RHIC and LHC energies, a parton scattering cross section of 1.5 mb is needed at both collision energies~\cite{Xu11,Xu11s} for the original AMPT model. For the improved AMPT model, the parton scattering cross sections of 3 mb at $\sqrt{s_{NN}}=200$ GeV and 2 mb at $\sqrt{s_{NN}}=2.76$ TeV are needed, respectively. Using the same event plane method as in the experimental analysis detailed in Appendix~\ref{flow}, the $p_T$ dependence of $v_2$ from AMPT model calculations can reproduce the STAR and the ATLAS data reasonably well, as shown in Fig.~\ref{v2pt}.

\subsection{Scaling relation of anisotropic flows at RHIC and LHC}

\begin{figure}[ht]
	\includegraphics[scale=0.15]{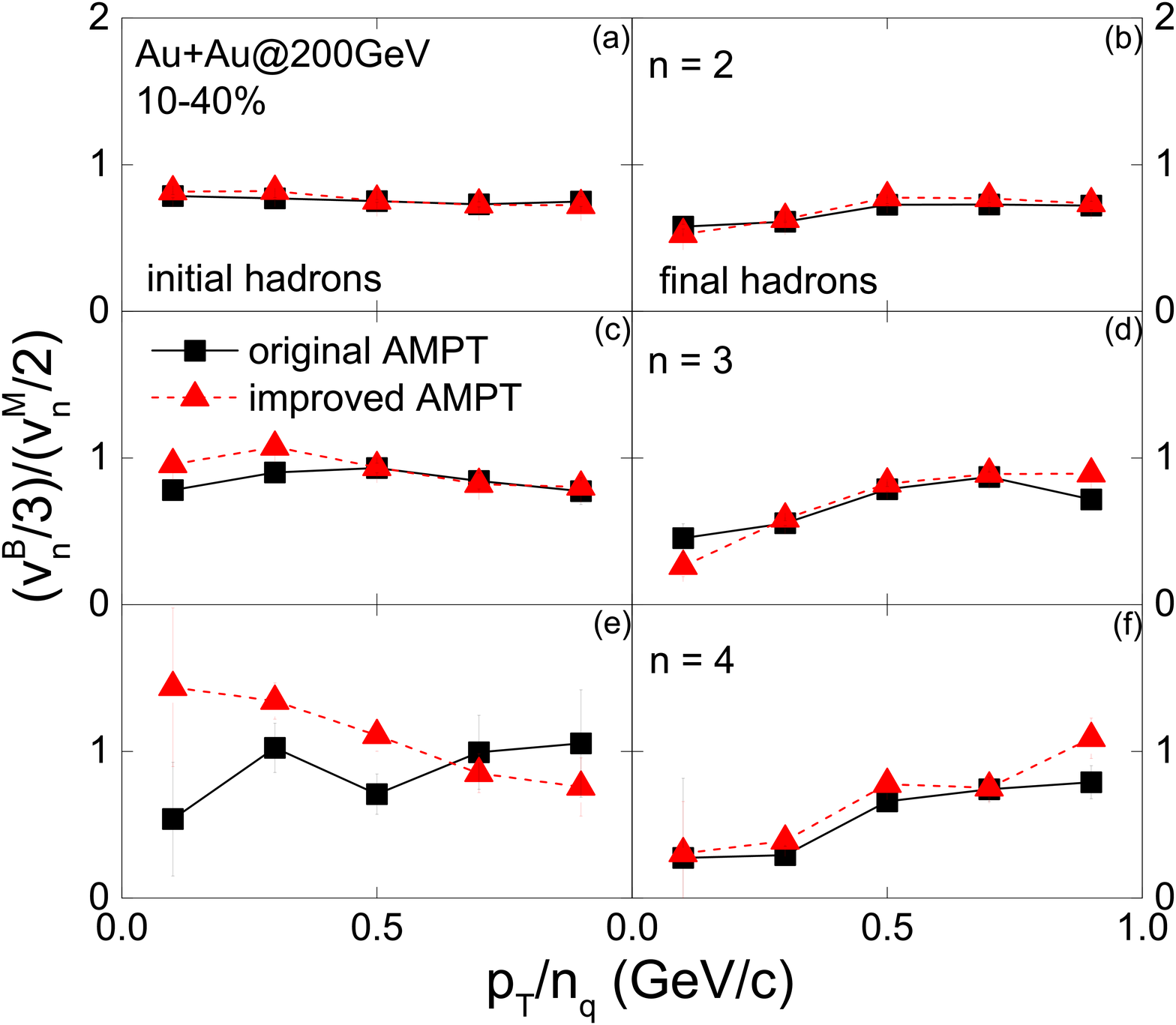}
	\caption{(Color online) Baryon-to-meson ratios of $v_{n} / n_{q}$ at the initial stage of the hadronic phase right after hadronization (left) and at the final stage of the hadronic phase after hadronic evolution (right) as a function of $p_{T} / n_{q}$ characterizing the NCQ scaling of anisotropic flows. Results from the original and the improved AMPT model in midcentral ($10 \%-40 \%$) Au+Au collisions at $\sqrt{s_{N N}}=200$ GeV are compared.}
	\label{200ncq}
\end{figure}

\begin{figure}[ht]
	\includegraphics[scale=0.15]{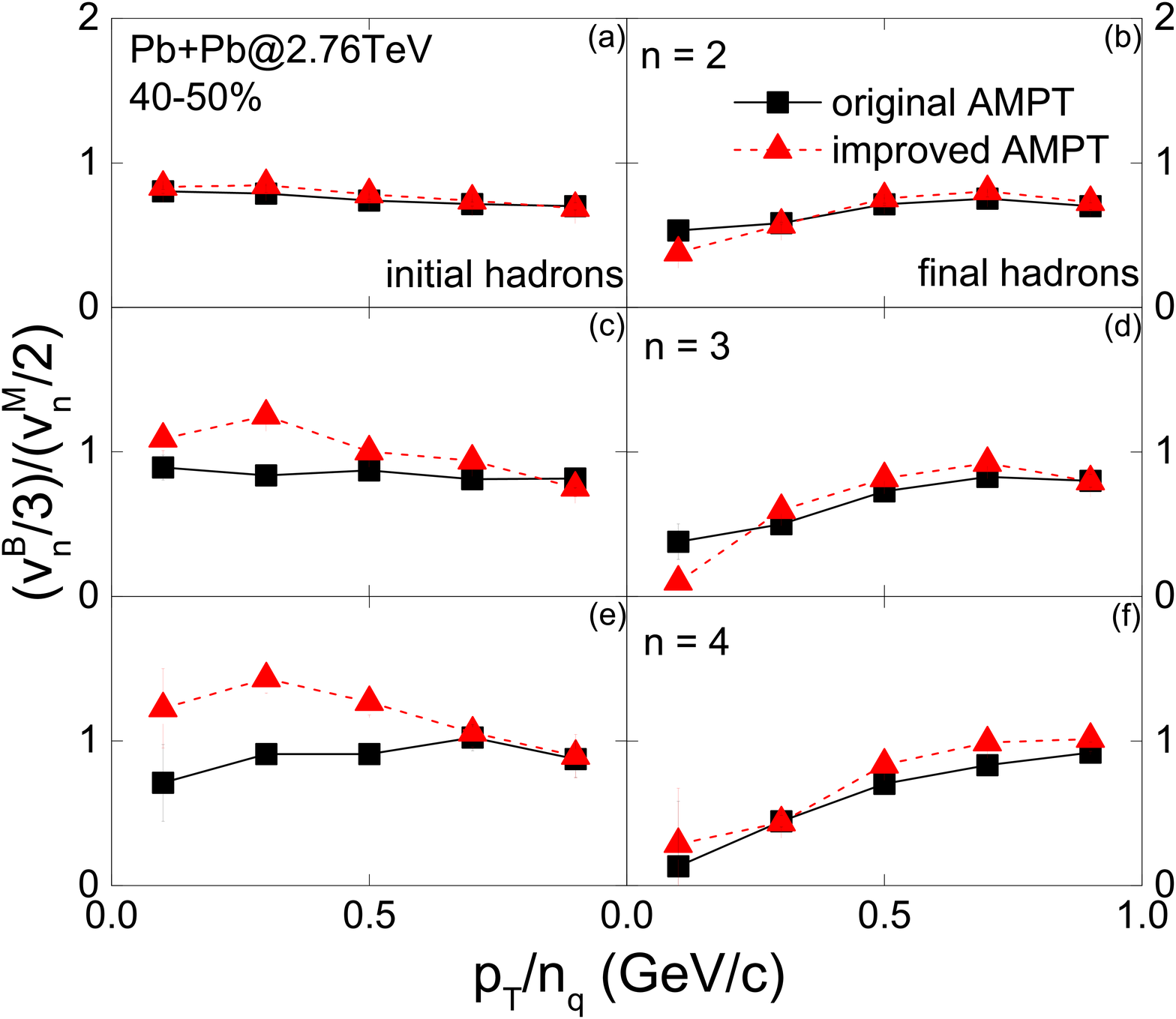}
	\caption{(Color online) Same as Fig.~\ref{200ncq} but in midcentral ($40 \%-50 \%$) Pb+Pb collisions at $\sqrt{s_{N N}}= 2.76$ TeV.}
	\label{2760ncq}
\end{figure}

In this subsection, we investigate the effects of the hadronization mechanism on the scaling relation of anisotropic flows in Au+Au collisions at $\sqrt{s_{NN}}=200$ GeV and Pb+Pb collisions at $\sqrt{s_{NN}}=2.76$ TeV. Theoretically, the NCQ scaling of $v_n$ can be obtained in the naive coalescence scenario by neglecting higher-order corrections if the relative distance in momentum space among valence partons is negligible~\cite{Mol03,Kol04}. The baryon-to-meson ratios of $v_{n} / n_{q}$ as a function of $p_{T} / n_{q}$ at top RHIC and LHC energy from different hadronization algorithms are compared in Figs.~\ref{200ncq} and \ref{2760ncq}, respectively. At both collision energies, elliptic flows ($v_2$), triangular flows ($v_3$), and quadratic flows ($v_4$) from both the original and the improved AMPT model satisfy the NCQ scaling law reasonably well, while the improved AMPT model seems to have systematically larger flows for baryons relative to mesons compared with the original AMPT model. After the hadronic evolution, the NCQ scaling seems to be violated to some extent, especially for low-$p_T$ hadrons that suffer from more collisions.

\begin{figure}[ht]
	\includegraphics[scale=0.3]{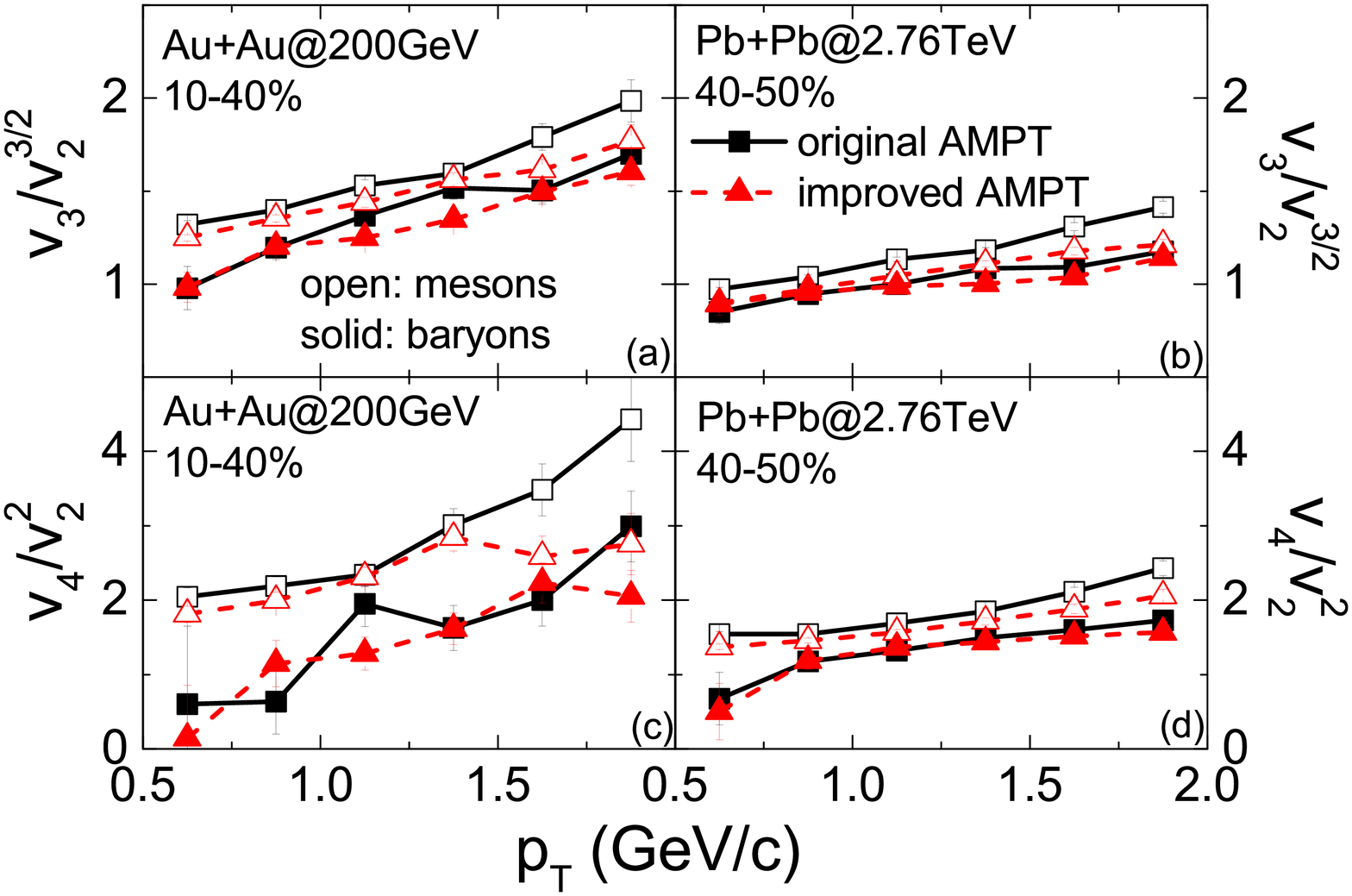}
	\caption{(Color online) Ratios of anisotropic flows for different orders as a function of the transverse momentum for final hadrons in midcentral ($10 \%-40 \%$) Au+Au collisions at $\sqrt{s_{N N}}=200$ GeV and midcentral ($40 \%-50 \%$) Pb+Pb collisions at $\sqrt{s_{N N}}= 2.76$ TeV from the original and the improved AMPT model.}
	\label{vnv2}
\end{figure}

The $v_n/v_2^{n/2}$ ratios for different order $n$ as a function of the transverse momentum for mesons and baryons at top RHIC and LHC energies from the original and the improved AMPT model are compared in Fig.~\ref{vnv2}. It is seen that the $v_n/v_2^{n/2}$ ratios are generally larger for mesons than for baryons, consistent with that observed in Ref.~\cite{Zha16}.  The improved AMPT model generally leads to smaller $v_n/v_2^{n/2}$ ratios and better scaling relations than the original AMPT model. This shows that the larger parton scattering cross section used in the improved AMPT model leads to the same $v_2$ but different $v_3$ and $v_4$ compared with the original AMPT model. The scaling relation as a function of $p_T$ is seen to be better satisfied at the LHC energy than at the top RHIC energy, presumedly due to more particles and violent collisions at higher energies. It is thus seen that the $v_n/v_2^{n/2}$ ratios are not only sensitive to the initial eccentricities~\cite{Lac10,Lac11b}, the viscous property of QGP~\cite{Gar12}, and the acoustic nature of anisotropic flows~\cite{Lac11a}, but depend on the hadronization mechanism as well.

\subsection{Directed flow at RHIC-BES}

\begin{figure}[ht]
	\includegraphics[scale=0.35]{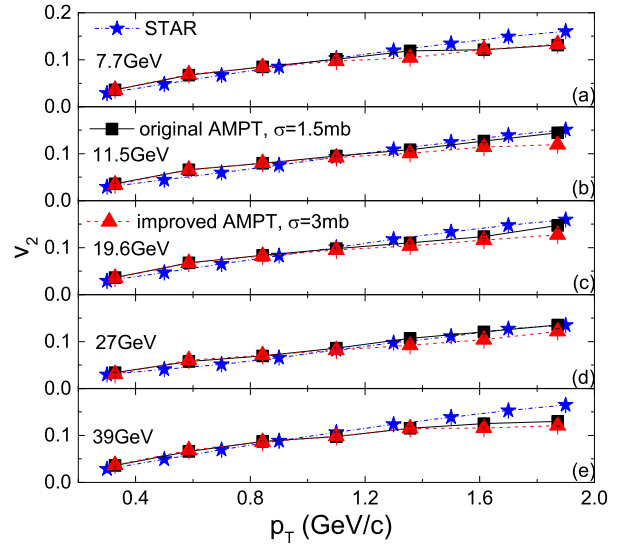}
	\caption{(Color online) Transverse momentum dependence of the charged hadron elliptic flow at midpseudorapidities in midcentral ($20 \%-30 \%$) Au+Au collisions at $\sqrt{s_{NN}}=$ 7.7, 11.5, 19.6, 27, and 39 GeV from the original and the improved AMPT model with the parton scattering cross sections of 1.5 mb and 3 mb, respectively, compared with the STAR data taken from Ref.~\cite{Ada12}.}
	\label{7739v2}
\end{figure}

In this subsection, we investigate the effects of the hadronization mechanism on the proton directed flow at RHIC-BES energies. We first reproduce the transverse momentum dependence of the charged particle elliptic flow at midpseudorapidities in midcentral Au+Au collisions at $\sqrt{s_{N N}}=7.7$ GeV to 39 GeV, as shown in Fig.~\ref{7739v2}. For the original and the improved AMPT model, the parton scattering cross sections of 1.5 mb and 3 mb are needed, respectively.

\begin{figure}[ht]
	\includegraphics[scale=0.3]{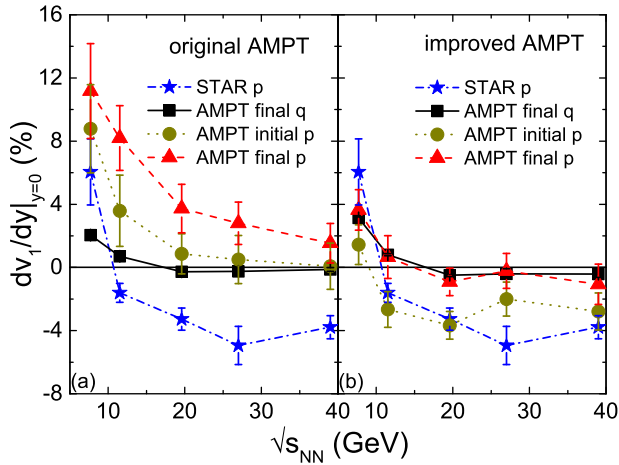}
	\caption{(Color online) Directed flow slope $d v_{1} / dy|_{y=0}$ of the final freeze-out quarks (final $q$), the initial protons (initial $p$) right after hadronization, and the final freeze-out protons (final $p$) versus the collision energy in midcentral ($20\%-30\%$) Au+Au collisions from the original (left) and the improved (right) AMPT model with the parton scattering cross sections of 1.5 mb and 3 mb, respectively. Results for final protons from the STAR Collaboration~\cite{Ada14} in midcentral ($10\%-40\%$) Au+Au collisions are also shown for comparison.} \label{proton}
\end{figure}

The slopes of the directed flows at midrapidities ($d v_{1} / dy|_{y=0}$) at different stages in midcentral Au+Au collisions are compared at different RHIC-BES energies in Fig.~\ref{proton}. For all results, the slopes are obtained from a cubic fit of the rapidity dependence of the directed flow, i.e., $v_{1}(y)=$ $F_{1} y+F_{3} y^{3}$. For both the original and the improved AMPT model using different parton scattering cross sections, the slopes of freeze-out parton directed flows change from positive to negative between 11.5 and 19.6 GeV. For the original AMPT model with spatial coalescence for partons, the negative slope of the directed flow is not preserved during the hadronization, as shown in Refs.~\cite{Guo17,Guo18}, and the slopes of the initial proton directed flow become positive at all collision energies. After hadronic rescatterings, the slope of the proton directed flow becomes even positively larger, inconsistent with the STAR data. For the improved AMPT model, the initial proton directed flow is more consistently converted from the freeze-out parton directed flow, and has a negative slope at the collision energy larger than 7.7 GeV. It is found that these directed flow slopes are qualitatively similar to those from the usual dynamical coalescence method based on the same freeze-out parton phase-space distribution. After the hadronic rescatterings, the slope of the proton directed flow becomes smaller but still negative at higher collision energies, qualitatively consistent with the STAR data. To reproduce quantitatively the STAR results of the $v_1$ slope requires handling more accurately the dynamics, while the significant effect of the hadronization mechanism on the directed flow has already been observed here.

\subsection{Relative particle yield ratios}

In this subsection, we investigate the effects of the hadronization mechanism on the relative particle yield ratios from $\sqrt{s_{NN}}=7.7$ GeV to 2.76 TeV. We first compare the meson-to-baryon yield ratio and the baryon-to-antibaryon yield ratio at midrapidities at RHIC-BES energies in Table~\ref{table1} and Table~\ref{table2}, respectively, for different scenarios. With the baryon chemical potential $\mu_B$ and the temperature $T$ at chemical freeze-out extracted from the STAR data~\cite{Ada17}, we are able to calculate the experimental meson-to-baryon yield ratio and baryon-to-antibaryon yield ratio according to the statistical model (see Appendix~\ref{statisticalmodels}). For the original AMPT model, there are two scenarios with different quark coalescence ordering, i.e., doing quark-antiquark coalescence for mesons first, or doing three-quark (three-antiquark) coalescence for baryons (antibaryons) first. For the improved AMPT model with the quark coalescence probability described by the Wigner function, there is no ambiguity of the coalescence ordering. Compared with results from the statistical model using the experimentally extracted chemical potentials and temperatures, none of the scenarios reproduces the meson-to-baryon and baryon-to-antibaryon yield ratios satisfactorily at all RHIC-BES energies. Typically, it is difficult to reproduce both the meson-to-baryon and baryon-to-antibaryon yield ratios correctly based on the same numbers of quarks and antiquarks, since enhancing the number of mesons generally reducing the numbers of baryons and antibaryons simultaneously, thus increasing both the meson-to-baryon and baryon-to-antibaryon yield ratios. Other mechanisms are needed to reproduce the relative particle yield ratios at RHIC-BES energies, such as modifying the initial quark components or introducing inelastic scatterings between partons.

\begin{table*}\small
	\centering
	\caption{The meson-to-baryon yield ratios at different RHIC-BES energies from the statistical model using the experimentally extracted baryon chemical potential $\mu_B$ and temperature $T$ at chemical freeze-out~\cite{Ada17}, from the original AMPT model with the hadronization for mesons before baryons and antibaryons (meson-baryon) or with the hadronization for mesons after baryons and antibaryons (baryon-meson), and from the improved AMPT model.}
	\begin{tabular}{|c|c|c|c|c|c|}
		\hline
		meson/baryon & 7.7 GeV & 11.5 GeV & 19.6GeV & 27 GeV & 39 GeV \\
		\hline
	    statistical model using exp $(\mu_B,T)$ & 2.24  & 3.59  & 5.91  & 7.20  & 8.01 \\
		\hline
		 original AMPT (meson-baryon) & $2.45 \pm 0.02$ & $4.28 \pm 0.02$ & $5.76 \pm 0.03$ & $6.62 \pm 0.04$ & $7.66 \pm 0.05$ \\
		\hline
		 original AMPT (baryon-meson) & $2.44 \pm 0.01$ & $4.22 \pm 0.02$ & $5.49 \pm 0.03$ & $6.07 \pm 0.04$ & $6.85 \pm 0.04$ \\
		\hline
		 improved AMPT & $2.19 \pm 0.01$ & $3.82 \pm 0.02$ & $5.16 \pm 0.03$ & $5.72 \pm 0.04$ & $6.18 \pm 0.04$ \\
		\hline

	\end{tabular}
	\label{table1}
\end{table*}

\begin{table*}\small
	\centering
	\caption{Same as Table \ref{table1} but for the baryon-to-antibaryon yield ratios.}
	\begin{tabular}{|c|c|c|c|c|c|}

\hline
		baryon/antibaryon & 7.7 GeV & 11.5 GeV & 19.6GeV & 27 GeV & 39 GeV \\
		\hline
		statistical model using exp $(\mu_B,T)$ & 200.65  & 38.04  & 9.68  & 5.31  & 3.48 \\
		\hline
		original AMPT (meson-baryon) & $193.34 \pm 12.83$ & $29.28 \pm 0.43$ & $9.64 \pm 0.10$ & $6.10 \pm 0.06$ & $3.95 \pm 0.03$ \\
		\hline
		original AMPT (baryon-meson) & $288.97 \pm 9.88$ & $35.54 \pm 0.59$ & $10.65 \pm 0.11$ & $6.56 \pm 0.06$ & $4.12 \pm 0.03$ \\
		\hline
		improved AMPT & $95.24 \pm 2.43$ & $28.36 \pm 0.51$ & $9.64 \pm 0.09$ & $5.54 \pm 0.05$ & $3.56 \pm 0.03$ \\
		\hline
	\end{tabular}
	\label{table2}
\end{table*}

\begin{figure}[ht]
	\includegraphics[scale=0.3]{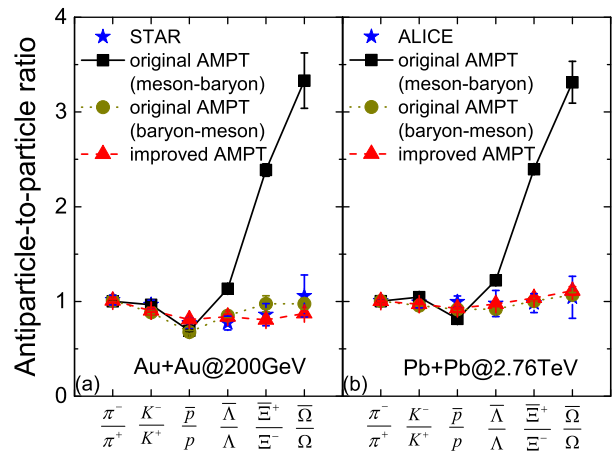}
	\caption{(Color online)  Antiparticle-to-particle yield ratios at midrapidities in central Au+Au collisions at $\sqrt{s_{NN}}=200$ GeV (left) and Pb+Pb collisions at $\sqrt{s_{NN}}=2.76$ TeV (right) from the original AMPT model with the hadronization for mesons before baryons and antibaryons (meson-baryon) or with the hadronization for mesons after baryons and antibaryons (baryon-meson), and from the improved AMPT model. Results from the STAR Collaboration~\cite{Ada07,Sui03,Abe09} and the ALICE Collaboration~\cite{Abe13,Sch15,Abe14} are also shown for comparison.} \label{ratio}
\end{figure}

Figure~\ref{ratio} compares the $\pi^{-}/\pi^{+}$, $K^{-}/K^{+}$, $\overline{p}/p$, $\overline{\Lambda}/\Lambda$, $\overline{\Xi}^{+}/\Xi^{-}$, and $\overline{\Omega}^{+} / \Omega^{-}$ yield ratios from the original and the improved AMPT with those from STAR and ALICE Collaborations, in central Au+Au collisions at $\sqrt{s_{NN}}=200$ GeV and Pb+Pb collisions at $\sqrt{s_{NN}}=2.76$ TeV, respectively. It is seen that the scenario of the original AMPT model, which does quark-antiquark coalescence first for mesons and then three-quark (three-antiquark) coalescence for baryons (antibaryons), overestimated significantly the $\overline{\Lambda}/\Lambda$, $\overline{\Xi}^{+}/\Xi^{-}$, and $\overline{\Omega}^{+} / \Omega^{-}$ yield ratios. This is understandable since in this scenario there are not too much choices for the combinations of strange baryons (antibaryons) after meson formation. This defect can be overcome by changing the coalescence ordering for mesons and baryons (antibaryons) in the original AMPT model. It is also remarkably seen that the improved AMPT model reproduces reasonably well these relative particle yield ratios at top RHIC and LHC energies. Although the results can be slightly dependent on the ordering of partons for baryon (antibaryon) coalescence, the difference turns out to be very small. These ratios can also be reproduced reasonably well by adjusting the coalescence parameter for meson and baryon formations in Ref.~\cite{He17}.

\subsection{Di-hadron correlations in p+p collisions}

\begin{figure}[ht]
	\includegraphics[scale=0.18]{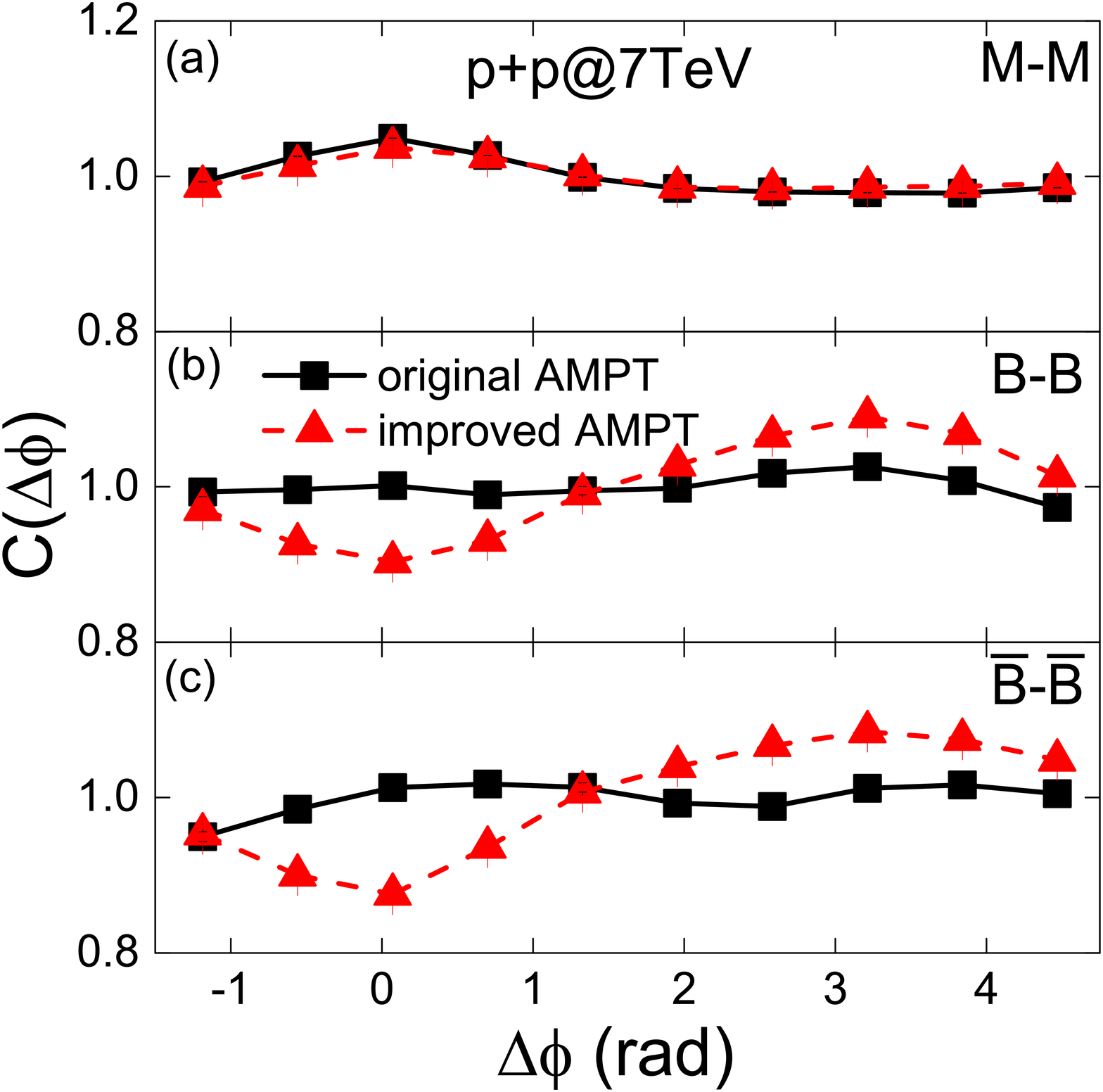}
	\caption{(Color online) Azimuthal angular dependence of the meson-meson (M-M) (a), baryon-baryon (B-B) (b), and antibaryon-antibaryon ($\rm\bar{B}$-$\rm\bar{B}$) (c) correlations in p+p collisions at $\sqrt{s_{NN}}=7$ TeV from the original and the improved AMPT model.} \label{correlation}
\end{figure}

As an illustration of the improved hadronization using Wigner functions as parton coalescence probabilities in the present study, we employ the parameter set B in Ref.~\cite{Xu11} and calculate the azimuthal angular dependence of di-hadron correlations for midrapidity hadrons in p+p collisions at $\sqrt{s_{NN}}=7$ TeV. The di-hadron correlation is defined as
\begin{equation}
C(\Delta \phi)=S(\Delta \phi)/B(\Delta \phi),
\end{equation}
where $\Delta \phi$ is the azimuthal angle difference between two hadrons of transverse momenta $p_T<2.5$ GeV/c, and $S(\Delta \phi)$ and $B(\Delta \phi)$ are the correlations in the same event and in mixed events, representing the signal and the background, respectively. As shown in Fig.~\ref{correlation}, both the original and the improved AMPT model lead to a near-side ($\Delta \phi \sim 0$) peak for the meson-meson correlation. However, the original AMPT model gives weak positive baryon-baryon and antibaryon-antibaryon correlations at the near side, while clear anti-correlation structures are observed from the improved AMPT model. The latter results are qualitatively consistent with the ALICE data, confirming that the near-side anti-correlation structure can be obtained with a more reasonable hadronization. Similar results and conclusions are obtained in Ref.~\cite{Zha18}.

\section{summary and outlook}
\label{summary}

Based on the string melting AMPT model, we have discussed the impact of the improved hadronization algorithm that favors parton combinations close in phase space. It was found that the improved AMPT model using this coalescence algorithm generally leads to a weaker elliptic flow from the same parton freeze-out phase-space distribution once parton combinations with similar momenta are favored. Thus, a larger parton scattering cross section is needed to reproduce the same experimental elliptic flow, compared with the original AMPT model using a spatial coalescence approach. Both the original and the improved AMPT model give reasonably well although slightly different scaling relations for anisotropic flows. Since the AMPT model with the improved quark coalescence algorithm preserves better the parton dynamics during the hadronization process, it shows a qualitatively correct collision energy dependence of the directed flow slope for protons. The relative particle yield ratios are also shown to be sensitive to the detailed hadronization algorithm, but other mechanisms are still needed to reproduce the experimental data especially at RHIC beam-energy-scan energies. The anti-correlation structures at the near side of baryon-baryon and antibaryon-antibaryon correlations are observed in p+p collisions from the improved AMPT model, but not in the original AMPT model.

The improved coalescence method developed in the present study can be further extended to the mix-event coalescence scenario in the spirit of the test-particle method when mean-field potentials for particles are incorporated, as in Ref.~\cite{Liu19}. Similar to the original spatial coalescence method, there are still partons left which are largely separated in phase space, since all partons are forced to be used up. These partons can be fragmented~\cite{And83} rather than coalesce into hadrons. On the other hand, the usual dynamical coalescence method using the perturbative treatment is generally used for rare particles (see, e.g., Ref.~\cite{Che03}). If one wants to compare in more detail results in the present study with those from the usual dynamical coalescence method, some modifications are needed in the later case by subtracting the formation probability once a parton is used to form a hadron in the later case. These go beyond the scope of the present study and need further investigations.

\begin{acknowledgments}

We thank Chen Zhong for maintaining the high-quality performance of the
computer facility. This work was supported by the National Natural Science
Foundation of China under Grant No. 11922514.

\end{acknowledgments}

\begin{appendix}

\section{Anisotropic flow analysis}	
\label{flow}

The momentum distribution of freeze-out hadrons can be expressed in terms of Fourier series for its azimuthal angular dependence as follows:
\begin{equation}
E \frac{d^{3} N}{d^{3} p}=\frac{1}{2 \pi} \frac{d^{2} N}{p_{T} d p_{T} d y}\left\{1+\sum_{n=1}^{\infty} 2 v_{n} \cos [n(\phi-\psi_{n})]\right\},
\end{equation}
where $\phi$ is the azimuthal angle of emitted particles, $p_{T}$ and $y$ are respectively the transverse momentum and the rapidity, the Fourier coefficient $v_{n}$ denotes the $n$th-order anisotropic flow, and $\psi_{n}$ denotes the corresponding event plane angle. In the following, we briefly review the event plane method in calculating the anisotropic flows. For more details, we refer the reader to Refs.~\cite{Pos98,Voloshin}.

The relationship  between the event flow vector $Q_{n}$ and the event plane angle $\psi_{n}$ for the $n$th-order harmonic term in the above distribution can be given by the following expressions
\begin{equation}
\begin{aligned}
Q_{n, x} &=Q_{n} \cos (n \psi_{n})=\sum_{i} \omega_{i} \cos (n \phi_{i}), \\ Q_{n, y} &=Q_{n} \sin (n \psi_{n})=\sum_{i} \omega_{i} \sin (n \phi_{i}),
\end{aligned}
\end{equation}
where the summations go over all particles used in the event plane calculation, and $\omega_{i}$ and $\phi_{i}$ are respectively the weight and the azimuthal angle for the $i$th particle. Here we set the weight factor $\omega_{i}$ as the transverse momentum of the $i$th particle. The event plane angle can thus be calculated from
\begin{equation}
\psi_{n}=\left[\operatorname{atan2} \frac{\sum_{i} \omega_{i} \sin (n \phi_{i})}{\sum_{i} \omega_{i} \cos (n \phi_{i})}\right] / n.
\end{equation}
With respect to the above event plane $\psi_{n}$, we can calculate the magnitude of the $n$th-order anisotropic flow $v_{n}^{obs}$ according to
\begin{equation}
v_{n}^{obs}(p_{T}, y)=\langle\cos [n(\phi_{i}-\psi_{n})]\rangle,
\end{equation}
where the angle bracket means averaging over particles in all events with their azimuthal angle $\phi_{i}$ for a given transverse momentum $p_{T}$ and rapidity ${y}$. To remove auto correlations, the contribution of the particle of interest from the total $Q_{n}$ vector must be subtracted, in order to obtain a $\psi_{n}$ uncorrelated with that particle. Since the finite multiplicity limits the estimation of the event plane angle, $v_{n}$ must be corrected by the event plane resolution for each ${n}$ given by
\begin{equation} \label{resolution}
\Re_{n}(\chi)= \frac{\sqrt{\pi}}{2} \chi \exp (-\chi^{2} / 2)[I_{0}(\chi^{2} / 2)+I_{1}(\chi^{2} / 2) ]
\end{equation}
with $\chi=v_{n}\sqrt{M}$, where ${M}$ is the multiplicity of particles, and $I_{0}$ and $I_{1}$ are the zeroth-order and first-order modified Bessel functions, respectively. In order to calculate the event plane resolution, the complete event is divided into two independent subevents with the same multiplicity of particles. Therefore, the resolution of subevents is just the square root of this correlation, i.e.,
\begin{equation}
\Re_{n}^{sub}=\sqrt{\langle\cos [n(\psi_{n}^{A}-\psi_{n}^{B})]\rangle},
\end{equation}
where $A$ and $B$ represent two subgroups of particles. In our calculation we divided particles within the pseudorapidity window, e.g. $|\eta|<1$, into two groups of forward and backward spheres with a gap of $|\Delta \eta|<0.1$. The complete event plane resolution can be obtained from
\begin{equation}
\Re_{n}^{full}=\Re(\sqrt{2} \chi_{sub}),
\end{equation}
where $\chi_{s u b}$ is inversely obtained from the subevent resolution $\Re_{n}^{sub}$ via Eq.~(\ref{resolution}). The final anisotropic flow is
\begin{equation}
v_{n}(p_{T}, y)=\frac{v_{n}^{obs}(p_{T}, y)}{\Re_{n}^{full}}.
\end{equation}

\section{Statistical model}
\label{statisticalmodels}

The chemical potential and the temperature at chemical freeze-out can be extracted experimentally from the relative particle yield based on the statistical model~\cite{sta,Cel06}. In the grand canonical ensemble, the density of the hadron species $i$ can be expressed as
\begin{equation}
\rho_{i}=g_{i} \int \frac{d^{3} p}{(2 \pi)^{3}} \frac{1}{e^{-(\sqrt{p^{2}+m_{i}^{2}}-\mu_{i})/T}\pm 1},
\end{equation}
where $\mu_{i}$, $m_{i}$, and $g_{i}$ are respectively the chemical potential, the mass, and the spin-isospin degeneracy of hadron species $i$, and $T$ is the temperature. The minus (plus) sign in the denominator is for mesons (baryons). The chemical potential of particle species $i$ can be expressed as
\begin{equation}
\mu_{i}=B_{i} \mu_{B}+Q_{i} \mu_{Q}+S_{i} \mu_{S},
\end{equation}
where $B_{i}$, $Q_{i}$, and $S_{i}$ are respectively the baryon number, the electric charge number, and the strangeness number of the hadron species $i$, and $\mu_{B}$, $\mu_{Q}$, and $\mu_{S}$ are the corresponding chemical potentials. Here we take the values of the chemical potentials and the temperatures at different RHIC-BES energies from Ref.~\cite{Ada17} extracted experimentally using the grand canonical ensemble ratio approach. The relative meson-to-baryon yield ratio and baryon-to-antibaryon yield ratio from the statistical model can thus be respectively expressed as
\begin{eqnarray}
\frac{N_M}{N_B} &=& \frac{\sum_i^M g_{i} \int d^{3} p \{\exp[-\sqrt{p^{2}+m_{i}^{2}}/T]- 1\}^{-1}}{\sum_i^B g_{i} \int d^{3} p \{\exp[-(\sqrt{p^{2}+m_{i}^{2}}-\mu_B)/T]+ 1\}^{-1}}, \notag\\
\frac{N_B}{N_{\bar{B}}} &=& \frac{\sum_i^B g_{i} \int d^{3} p \{\exp[-(\sqrt{p^{2}+m_{i}^{2}}-\mu_B)/T]+ 1\}^{-1}}{\sum_i^B g_{i} \int d^{3} p \{\exp[-(\sqrt{p^{2}+m_{i}^{2}}+\mu_B)/T]+ 1\}^{-1}}.\notag
\end{eqnarray}
The above ratios representing the experimental results are compared with those from the AMPT model.

\end{appendix}

\end{document}